\documentstyle[epsf,psfig,referee]{mn}

\begin{document}

\title[Anglo-Australian Planet Search]
{An exoplanet in orbit around $\tau^1$~Gruis}

\author[H. R. A. Jones et al.]{
\parbox[t]{\textwidth}{Hugh R. A. Jones$^1$, R. Paul Butler$^2$,  
Chris G. Tinney$^3$, Geoffrey W. Marcy$^4$,  
Alan J. Penny$^5$, Chris McCarthy$^2$,
Brad D. Carter$^6$} \\
\vspace*{6pt} \\
$^1$Astrophysics Research Institute, Liverpool John Moores University,
Egerton Wharf, Birkenhead CH41 1LD, UK\\
$^2$Carnegie Institution of Washington,Department of Terrestrial Magnetism,
5241 Broad Branch Rd NW, Washington, DC 20015-1305, USA \\
$^3$Anglo-Australian Observatory, PO Box 296, Epping. 1710, Australia\\  
$^4$Department of Astronomy, University of California, Berkeley, CA, 94720\\
$^5$Rutherford Appleton Laboratory, Chilton, Didcot, Oxon OX11 0QX, UK\\
$^6$Faculty of Sciences,  University of Southern Queensland, Toowoomba, 
QLD 4350, Australia\\
}

\date{September 2002}

\maketitle

\label{firstpage}

\begin{abstract}

We report the detection of a new candidate exoplanet around the metal-rich
star $\tau^1$~Gruis. With M~sin~$i$~=~1.23$\pm$0.18 M$_{\rm JUP}$,
a period of 1326$\pm$300~d and an orbit with an
eccentricity of 0.14$\pm$0.14 it adds to the growing population of
long period exoplanets with near-circular orbits. This population
now comprises more than 20\% of known exoplanets.

When the companion to $\tau^1$~Gruis is plotted together with 
all exoplanets found by
the Anglo-Australian Planet Search and other radial velocity searches 
we find evidence for a peak in the number of short-period exoplanets,
followed by a minimum of planets between around 7 and 50 days and
then an apparent rise in the number of planets per unit 
radius that seems to set in by a hundred days, 
indicating more planets farther from the host star.
This is very different from the gaussian-like 
period distribution found for stellar companions. 
This lends support to the idea that once a clearing in 
the inner protoplanetary disk develops, it halts the 
inward migration of planets. In particular, the smooth 
distribution of exoplanets arising from planetary migration
through a disk is altered by an accumulation of exoplanets
at the point where the disk has been cleared out.

\end{abstract}

\begin{keywords}
planetary systems - stars: individual (HD216435), 
brown dwarfs
\end{keywords}

\section{Introduction}
The Anglo-Australian Planet Search (AAPS) is a long-term planet
detection programme which aims to perform exoplanet 
detection and measurement at the highest possible precision.
Together with programmes using similar techniques on the
Lick 3\,m and Keck I 10\,m telescopes (Fischer et al. 2001; Vogt et al. 2000),
it provides all-sky planet search coverage for inactive F, G, K and M dwarfs
down to a magnitude limit of V=7.5. So far the AAPS has  
has published data for 17 exoplanets. (Tinney et al. 2001; 
Butler et al. 2001; Butler et al. 2002a; Jones et al. 2002a,b;
Tinney et al. 2002a,b).

The AAPS is carried out on the 3.9m
Anglo-Australian Telescope (AAT) using the University College London Echelle
Spectrograph (UCLES), operated in its 31 lines/mm mode
together with an I$_{2}$ absorption cell.
UCLES now uses the AAO's EEV 2048$\times$4096 13.5$\mu$m pixel CCD,
which provides excellent quantum efficiency across the 500--620~nm
I$_2$ absorption line region.
Despite this search taking place on a common-user telescope
with frequent changes of instrument, we achieve a 3~m~s$^{-1}$ precision
down to the V~=~7.5 magnitude limit of the survey (Butler et al. 2001;
fig. 1, Jones et al. 2002a).

Our target sample, which we have observed since 1998, is given in
Jones et al. (2002b). It includes 178 late (IV-V) F, G and K
stars with declinations below $\sim -20^\circ$ and is complete
to V$<$7.5. We also observe sub-samples of 16 metal-rich
([Fe/H]$>$0.3) stars with V$<$9.5 and 7 M dwarfs with V$<$7.5 and
declinations below $\sim -20^\circ$. The sample is being increased
to around 300 solar-type stars to be complete to a magnitude limit of V=8.
Where age/activity information is
available from log~$R$'(HK) indices (Henry et al. 1996; Tinney et al. 2002c)
we require target stars to have log~$R$'(HK) $<$ --4.5
corresponding to
ages greater than 3 Gyr. Stars with known stellar companions within
2 arcsec are removed from the observing list, as it is operationally difficult
to get an uncontaminated spectrum of a star with a nearby companion.
Spectroscopic binaries discovered during the programme have also been
removed and will be reported elsewhere (Blundell et al., in preparation).
Otherwise there is no bias against observing multiple stars. The programme
is also not expected to have any bias against brown dwarf companions.
The observing and data processing procedures
follow those described by Butler et al. (1996, 2001).

\section{Stellar Characteristics of $\tau^1$~Gruis}

The Bright Star Catalog
assigns $\tau^1$~Gruis a spectral type of G0V, compared to the
Hipparcos spectral type of G3IV.
Its parallax of 30.0$\pm$0.7\,mas (ESA 1997) together
with a V magnitude of 6.03 imples an absolute magnitude 
of M$_V$~= 3.42$\pm$0.03 and M$_{\rm bol}$=3.20$\pm$0.05 (Cayrel
et al. 1997). This absolute magntidue puts $\tau^1$~Gruis
a magnitude above the main sequence and 
explains the discrepancy between the literature assigned 
spectral types G0V and G3IV.

Figure 1 shows the Ca $\sc{II}$ H line for $\tau^1$~Gruis
(HD~216435, HIP113044, HR8700) and indicates it is
chromospherically inactive, confirming the activity index
log~$R$'(HK)~=~--5.00 found by Henry et al. (1996).
Furthermore there is no evidence for significant photometric
variability in the 121 measurements made by the Hipparcos satellite.
Combining Hipparcos astrometry of $\tau^1$~Gruis with its SIMBAD radial velocity
yields a space velocity with respect to the local standard of rest:
U, V, W = --27.5, --21.7, --10.5. Its inferred age is 5 Gyr(Gonalez 1999).
Favata, Micela \& Sciortino (1996) report an equivalent
width Li detection of 70~m\AA\ equating to an abundance of
lithium N(Li) = 2.44, consistent with other similar 
metal-rich sub-giants (Randich et al. 1999).
di Benedetto (1998) has calculated
the effective temperature of $\tau^1$~Gruis to be 5943$\pm$60 K as part of 
a substantial programme to apply the infrared flux method and angular 
diameters from interferometry experiments to ISO standard stars 
with V-K measurements.
Like many of the stars found to have extra-solar planets, 
$\tau^1$~Gruis is metal
rich with [Fe/H] derived from high resolution spectroscopic analysis
of Fe lines is +0.15$\pm0.04$ (Favata, Micela \& Sciortino 1997).
Interpolation
between the tracks of Fuhrmann, Pfeiffer \& Bernkopf (1998)
and Girardi et al. (2000) indicates a mass of 1.25$\pm$0.10~M$_\odot$.



\section{Orbital Solution for $\tau^1$~Gruis}

The 40 Doppler velocity measurements of $\tau^1$~Gruis, obtained
between 1998 August and 2002 August, are shown graphically in Figure 2
and listed in Table 1. The third column labelled uncertainty
is the velocity uncertainty produced by our least-squares
fitting.
This uncertainty includes the effects of photon-counting uncertainties,
residual errors in the spectrograph PSF model, and variation in
the underlying spectrum between the template and iodine epochs. All
velocities are measured relative to the zero-point defined by
the template observation. Only observations where the uncertainty
is less than twice the median uncertainty are listed.

The data are well-fit by a Keplerian curve which 
yields an orbital period of $1326\pm300$~d, a velocity amplitude
of $20\pm2$~m~s$^{-1}$, and
an eccentricity of $0.14\pm0.14$. The minimum (M~sin~$i$) mass of the
planet is $1.2\pm0.1$~M$_{\rm JUP}$, and the semi-major axis is
$2.6\pm0.6$~au. The RMS to the Keplerian fit is 6.92~m~s$^{-1}$,
yielding a reduced chi-squared of 1.24. 
Since $\tau^1$~Gruis is relatively bright (V=6.03) and inactive, the 
measured RMS seems a little high. We have investigated our data
and find that the internal velocity errors correlate
well with the number of photons per pixel.
The velocity errors vary nearly as 1/$\sqrt{(\rm photons)}$,
as one would expect.  We consider that there is some constant
error of around $\sim$3 m~s$^{\rm -1}$, probably caused by 
inadequate S/N in our template measurement. We also note that 
at the spectral type of $\tau^1$~Gruis, 
G0V, our precision is limited by the smaller equivalent widths
of stellar lines relative to later spectral types. Thus with our 
observing exposures adjusted to give S/N=200 per exposure our 
precision is probably limited
to around 4 rather than 3 m~s$^{\rm -1}$. The lack of any observed
chromospheric activity or photometric variations gives us confidence
that the radial velocity signature arises from an exoplanet rather
than from long-period starspots or chromospherically active regions.  
The properties of the candidate 
extra-solar planet in orbit around $\tau^1$~Gruis are summarised
in Table 2.

\section{Discussion}

The companion to $\tau^1$~Gruis announced here serves to further reinforce
the predominantly metal-rich nature of stars with exoplanets. It also adds to the growing population of long period exoplanets with near-circular orbits.
Now more than 20\% of exoplanets have orbital parameters within those of the Solar System. 
It is notable that as the Anglo-Australian Planet Search becomes sensitive to longer periods, we are continuing to find objects with longer periods, but remain limited by our first epoch observations.
$\tau^1$~Gruis is a pleasing example. Within the errors its velocity 
amplitude is nearly as low as any long-period single exoplanet announced by radial velocity searches and the error on
its period is dominated by our first
epoch observation. Thus the detection of an exoplanet around $\tau^1$~Gruis 
together with our long-term stable stars (e.g., Butler et al. 2001) 
gives us confidence in the stability of our search as we 
move to longer periods and the possibility of detecting 
Jupiter analogues.

The radial velocity signal we measure for $\tau^1$~Gruis suggests a planet with
a minimum mass around that of Jupiter. $\tau^1$~Gruis~b becomes the fifth exoplanet to be found with a mass around that of Jupiter with a period of greater than three years and indicates that radial velocity surveys now have significant sensitivity to Jupiter mass planets out to relatively large periods.

 It is thus intriguing to look at the period distribution of exoplanets found by the AAPS and other radial velocity searches.
In Butler et al. (2002b), we plotted a histogram of semimajor axes for exoplanets from the Lick, Keck and AAT searches. This showed a relatively large number of exoplanets at very short orbits and a tail of objects with longer orbits.
The detection of long period planets and long-term stable stars indicated that the peak at short periods was a real feature. 
Two years later, with twice as many exoplanets known, 
Figure 3 shows the exoplanets that have been announced based on exoplanets.org
by 2002 August 21. 
The bulk of known exoplanets now lie at relatively large periods.
Although the AAPS has been operating for less time than other successful 
searches, Figure 3 also shows that the exoplanets published by the AAPS 
are dominated by companions at longer periods such as $\tau^1$~Gruis~b. 
The top part of Figure 3 shows a peak at shorter periods together with 
a substantial fraction at longer periods. Interestingly there appears 
to be a gap in the distribution between periods of around 7 and 50 days. 
The evidence for this gap is relatively poor when considering the AAT 
planets alone though striking when all exoplanets are considered. 

The relative lack of exoplanet candidates from around 0.2 to 0.6 AU was noted by Cummings, Marcy \& Butler (1999) and Butler et al. (2002b) and is also evident in fig. 2 of Heacox et al.(1999), fig. 5 of Rabachnik \& Tremaine 2001, fig. 4 of Lineweaver \& Grether (2002) and fig. 7 of Armitage et al. (2002). 
Armitage et al. interpret this feature as a slight excess of exoplanets at the shortest periods and attribute it to the completeness of radial velocity surveys falling off toward longer periods. However, the observed period distribution actually appears to rise toward longer periods where the incompleteness of the radial velocity surveys falls rapidly. To allow for this incompleteness introduced by including lots of low-mass short-period exoplanets we follow Armitage et al. (2002) and consider planets in a restricted mass and period range where the surveys can be judged to be more complete. Following the analysis of Cummings et al. (1999), Armitage et al. consider the known exoplanets to be complete in the mass range 0.6--10 M$_{\rm JUP}$ sin $i$ for periods of less than 3 AU. In the middle plot of Figure 3 we show the period distribution for a 0.6--10 M$_{\rm JUP}$ mass cut off as well as all announced planets. The removal of the lowest mass exoplanets reduces the peak of very short period planets, however, the peak at short periods remains an order of magnitude higher than would be expected from an extension of counts at longer periods. We have also looked at the CORALIE, Keck and Lick surveys in the same manner as for the AAPS. Despite different radial velocity surveys operating with different samples, sensitivities, instruments, scheduling, strategies and techniques we do find evidence for the gap in each of the major surveys. As mentioned above, this gap is evident in a number of works by other authors though is relatively less pronounced because of the substantially smaller number of exoplanets announced when those plots were made. The pronounced nature of this peak leads us to consider Fig. 4 to show evidence for two (or more) populations of exoplanets. That is, a population of exoplanets spanning a small range of short periods (3-7 days) and an separate population increasing with number towards larger periods.

From the lower part of Figure 3, it can be seen that there is no evidence for such a gap in the stellar binary distribution. The stellar companion period distribution plotted was determined by Duquennoy \& Mayor (1991) using the same general radial velocity method to discover binary stars as used to discover the exoplanets. Duquennoy \& Mayor find it necessary to make corrections to the stellar binary period distribution for incompleteness at longer periods, however, no such corrections are necessary for shorter periods and they find no gap in short-period stellar binaries. Overall Duquennoy \& Mayor find that the period distribution of stellar binaries is well fit by a gaussian. Since stellar companions are expected to form via large-scale gravitational instabilities in collapsing cloud fragments or massive disks, whereas planets are expected to form by accretion in dissipative circumstellar
disks it is not surprising that stellar and planetary companions should have different period distributions. 

The period distribution for exoplanets has been investigated a number of times as the number of radial velocity exoplanets has grown. The existence of a peak at very short periods has been an important motivation in the development of migration theories for exoplanets (Lin et al. 1999; Murray et al. 1998; Trilling et al. 1998; Ward 1997; Armitage et al. 2002; Trilling, Lunine \& Benz 2002). The trend toward finding an increasing number of exoplanets with large orbital separation runs counter to the selection effects inherent in radial velocity searches and has been well reproduced by migration theories (Armitage et al. 2002;
Trilling et al. 2002). 
Whilst selection effects start to play an increasingly important role beyond around few hundred days (e.g. Duquennoy \& Mayor 1991; Cummings et al. 1999; 
Butler et al. 2002b), we do not consider them to be significant between
7 and 50 days and thus consider the 'gap' in exoplanet periods  
to be a feature of the period distribution not currently predicted by migration theories. It is interesting to speculate on the origin of the possible gap
in the exoplanet period distribution. 

The onset of the stellar wind in young stars and the magnetic clearing 
of a hole at the centre of the disk will lead to the evacuation of the 
circumstellar disk and prevent migration of planets. This is expected to happen sooner in stars of higher mass and suggests the exoplanets of stars with higher mass will lie at greater radii. 
So far the range of stellar masses yielding significant numbers 
of exoplanets is rather small and we find no clear difference in 
exoplanet properties for stars of different mass.
Even without such evidence, migration theory does provide an attractive 
explanation for a range of exoplanet properties. Migration theory
can already reasonably explain the
progressively larger number of exoplanets at larger radii
and with the inclusion of
appropriate stopping mechanisms (e.g.  Lin, Bodenheimer \& Richardson 1996; 
Kuchner \& Lecar 2002) may also be able to consistently produce the peak
in the period distribution of short period planets.

\section*{Acknowledgments}
The Anglo-Australian Planet Search team would like to thank
the support of the Director of the AAO, Dr Brian Boyle,
and the superb technical support which has been received throughout
the programme from AAT staff -- in particular F.Freeman, D.James, S.Lee, J.Pogson, R.Patterson, D.Stafford and J.Stevenson.
We gratefully
acknowledge the UK and Australian government support of the
Anglo-Australian Telescope through their PPARC and DETYA funding
(HRAJ, AJP, CGT); NASA grant NAG5-8299 \& NSF grant
AST95-20443 (GWM); NSF grant AST-9988087 (RPB); and
Sun Microsystems.  This research has made use of the SIMBAD
database, operated at CDS, Strasbourg, France

\begin{figure}
\psfig{file=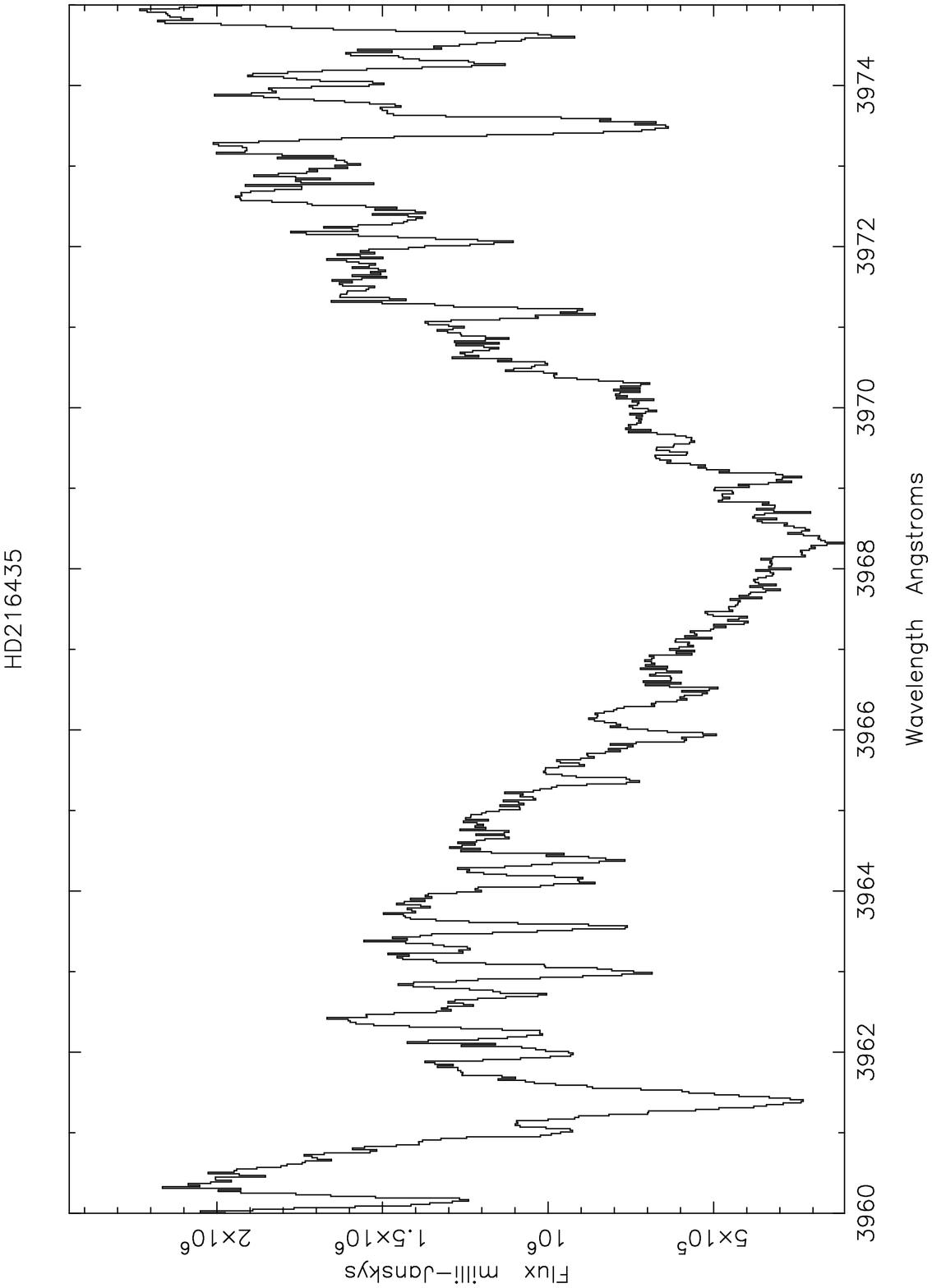,height=8.5in}
\caption{The figure shows the Ca$\sc{II}$ H line core in $\tau^1$~Gruis.
No emission is evident confirming the low activity index,
log~$R$'(HK)~=~--5.00, measured by Henry et al.
(1996). The activity of the entire AAPS sample will be assessed
in forthcoming papers by Tinney et al. (in preparation) and 
Blundell et al. (in preparation).}
\label{hd39091}
\end{figure}

\begin{figure}
\psfig{file=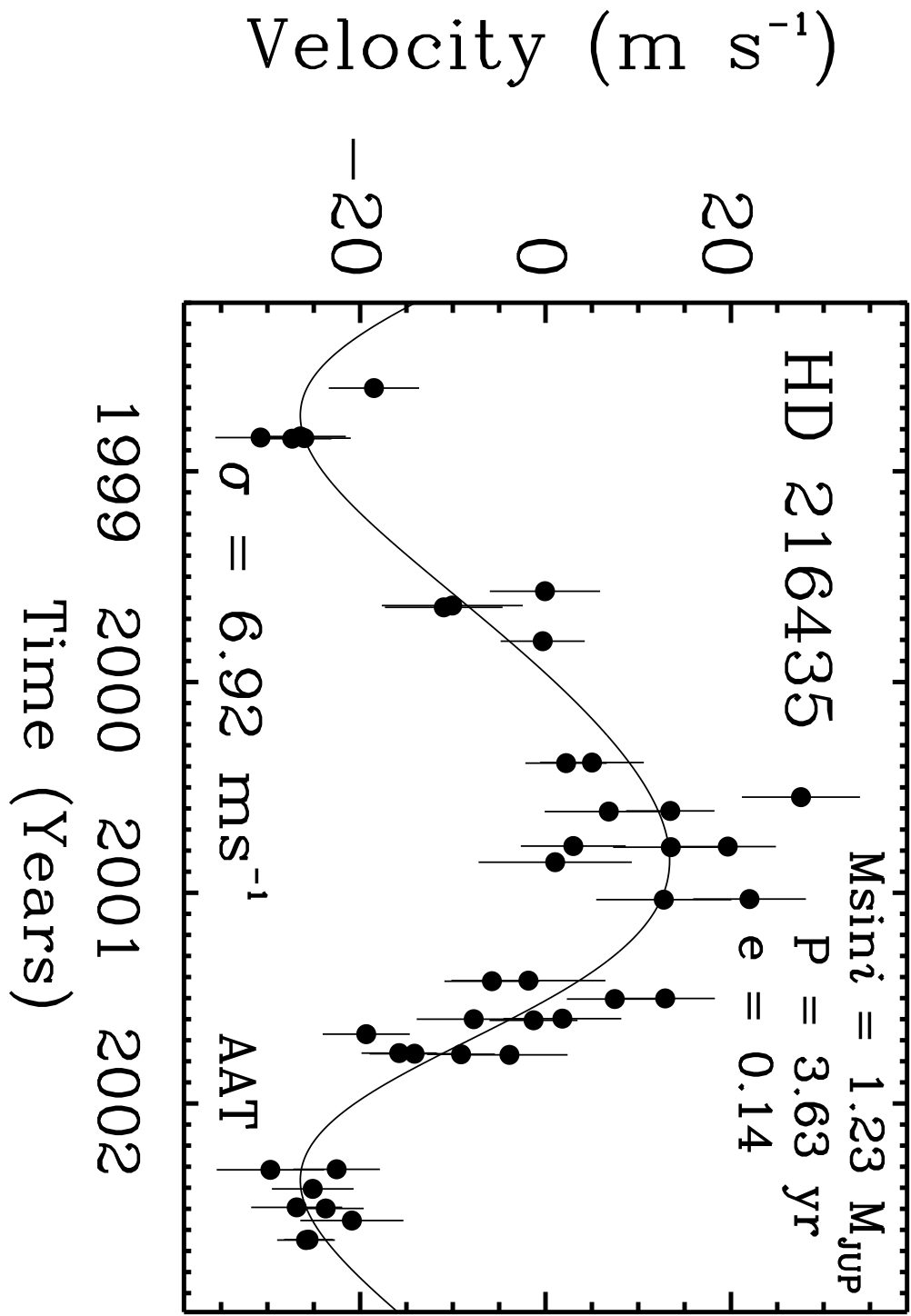,height=8.5in}
 \caption{Doppler velocities obtained for $\tau^1$~Gruis from 1998 August to
2002 August.
The solid line is a best fit Keplerian with the parameters shown
in Table 1.
The rms of the velocities about the fit is 6.83~m~s$^{\rm -1}$.
Assuming 1.25~M$\odot$ for the primary,
the minimum (M~sin~$i$) mass of the companion is 1.23~M$_{\rm JUP}$ and
the semimajor axis is 2.6~au.}
\label{hd2039}
\end{figure}

\begin{figure}
\psfig{file=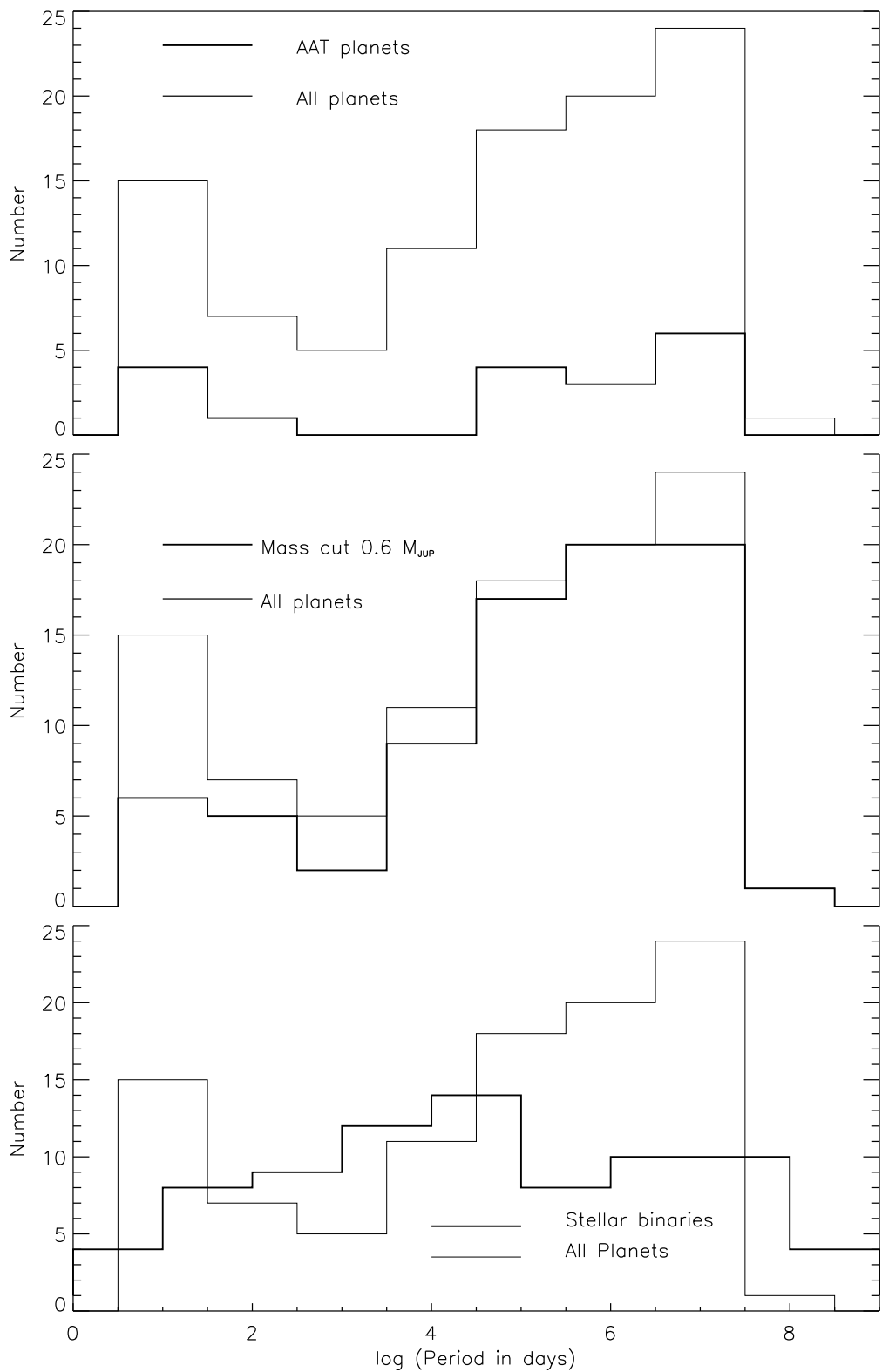,height=8.5in}
 \caption{The number of exoplanets discovered within natural 
logarithm period bins is shown. The top part of the figure compares 
all radial velocity planets announced based
on exoplanets.org/almanacframe.html (2002 August 21) with those 
published by the AAT. This includes planets in the AAT sample that were
first published by other planet search projects.
The middle part of the figure compares the period distributions of 
all planets to those with masses less 
than 10 M$_{\rm JUP}$ sin $i$ and greater 
than 0.6 M$_{\rm JUP}$ sin $i$. The bottom part 
compares all published planets with 
the stellar binaries as found by Duquennoy \& Mayor (1991). }
\label{}
\end{figure}

\newpage
\begin{table}
 \centering
 \begin{minipage}{140mm}

  \caption{Velocities for $\tau^1$~Gruis. Julian Dates (JD) are heliocentric.
Radial Velocities(RV) are heliocentric but have an arbitrary 
zero-point determined by the radial velocity of the template.}
  \begin{tabular}{@{}lrc@{}}
JD & RV & Error \\

-2451000   &  m~s$^{-1}$ & m~s$^{-1}$ \\
    34.2105  &    -9.5  &  4.9 \\
   118.0436  &   -17.4  &  4.9 \\
   119.9400  &   -21.7  &  4.9 \\
   120.9997  &   -17.0  &  5.0 \\
   121.9209  &   -18.3  &  4.2 \\
   386.3182  &     8.9  &  5.9 \\
   411.1493  &    -1.1  &  7.6 \\
   414.2635  &    -2.0  &  6.3 \\
   472.9492  &     8.7  &  4.5 \\
   683.3180  &    14.0  &  5.6 \\
   684.3245  &    11.2  &  4.4 \\
   743.2420  &    36.6  &  6.4 \\
   767.1997  &    22.5  &  4.8 \\
   768.2187  &    15.8  &  6.9 \\
   828.0383  &    12.0  &  5.7 \\
   828.9589  &    28.7  &  5.2 \\
   829.9527  &    22.5  &  6.2 \\
   856.0436  &    10.0  &  8.3 \\
   919.9251  &    31.0  &  6.1 \\
   920.9303  &    21.8  &  7.3 \\
  1061.2803  &     7.2  &  8.3 \\
  1062.3446  &     3.2  &  5.1 \\
  1092.2145  &    21.9  &  5.4 \\
  1093.2415  &    16.5  &  5.2 \\
  1127.1922  &    10.8  &  6.4 \\
  1128.1475  &     1.3  &  6.1 \\
  1130.1052  &     7.7  &  4.7 \\
  1154.0982  &   -10.3  &  4.7 \\
  1186.9518  &    -6.8  &  4.1 \\
  1188.0403  &    -5.1  &  4.8 \\
  1188.9741  &    -0.1  &  3.6 \\
  1189.9808  &     5.1  &  6.3 \\
  1388.3132  &   -13.5  &  4.7 \\
  1389.3008  &   -20.7  &  5.8 \\
  1422.3045  &   -16.1  &  4.4 \\
  1454.3375  &   -17.8  &  4.9 \\
  1456.2765  &   -14.7  &  4.1 \\
  1477.1981  &   -11.9  &  5.6 \\
  1510.2436  &   -16.6  &  2.6 \\
  1511.0242  &   -16.8  &  3.1 \\
\end{tabular}
\end{minipage}
\end{table}

\begin{table}
 \centering
 \begin{minipage}{140mm}
  \caption{Orbital parameters for the companion to $\tau^1$~Gruis.}
  \begin{tabular}{@{}lrrr@{}}
Parameter & $\tau^1$~Gruis~b \\
Orbital Period (d) & 1326$\pm$300 \\
Eccentricity & 0.14$\pm$0.14 \\
$\omega$ ($\deg$)& 115$\pm$60 \\
Velocity amplitude K (m~s$^{-1}$ & 20$\pm$2 \\
Periastron Time (JD)  & 50894$\pm$300\\
M~sin~$i$ (M$_{\rm JUP}$) & 1.23$\pm$0.18 \\
a (AU) & 2.6$\pm$0.6 \\
RMS to Fit & 6.92 \\
\end{tabular}
\end{minipage}
\end{table}

\label{lastpage}
\end{document}